# Neural-Network-Enhanced Metalens Camera for High-Definition, Dynamic Imaging in the Long-Wave Infrared Spectrum

Jing-yang Wei,[1,†] Hao Huang,[1,†] Xin Zhang,[1] De-Mao Ye,[1] Yi Li,[1] Le Wang,[1] Yao-guang Ma,[2,*] and Yang-hui Li [1,*]

[1]College of Optical and Electronic Technology, China Jiliang University, Hangzhou 310018, China
[2]State Key Laboratory of Extreme Photonics and Instrumentation, College of Optical Science and Engineering; Intelligent Optics and Photonics Research Center, Jiaxing Research Institute; ZJU–Hangzhou Global Scientific and Technological Innovation Center; International Research, Center for Advanced Photonics, Hangzhou, Zhejiang 310013, China
[†]These authors contributed equally to this work.
*Correspondence should be addressed to Yao-guang Ma (mayaoguang@zju.edu.cn) and Yang-hui Li (lyh@cjlu.edu.cn).

**Abstract:** To provide a light weight and cost-effective solution for the long-wave infrared imaging using a singlet, we develop a camera by integrating a High-Frequency-Enhancing Cycle-GAN neural network into a metalens imaging system. The High-Frequency-Enhancing Cycle-GAN improves the quality of the original metalens images by addressing inherent frequency loss introduced by the metalens. In addition to the bidirectional cyclic generative adversarial network, it incorporates a high-frequency adversarial learning module. This module utilizes wavelet transform to extract high-frequency components, and then establishes a high-frequency feedback loop. It enables the generator to enhance the camera outputs by integrating adversarial feedback from the high-frequency discriminator. This ensures that the generator adheres to the constraints imposed by the high-frequency adversarial loss, thereby effectively recovering the camera's frequency loss. This recovery guarantees high-fidelity image output from the camera, facilitating smooth video production. Our camera is capable of achieving dynamic imaging at 125 frames per second with an End Point Error value of 12.58. We also achieve 0.42 for Fréchet Inception Distance, 30.62 for Peak Signal to Noise Ratio, and 0.69 for Structural Similarity in the recorded videos.

## Introduction

Traditional long-wave infrared (LWIR) optical systems use bulky, multi-element lenses to correct optical aberrations and achieve high-quality imaging. However, this reliance on multiple elements poses challenges for miniaturization in applications where compactness is essential. Metalenses, utilizing flat surfaces to manipulate light, offer a compact and lightweight alternative, enabling the creation of thinner and lighter optical systems [1]. This expands their potential applications in various fields such as imaging, spectroscopy, and communications [2-4]. However, the optical imaging performance of metalens is inevitably affected by aberrations. The chromatic aberrations originating from the subwavelength nature of metalenses pose a significant challenge that hinders the widespread adoption of metasurface lenses [5-11]. Besides, the image quality can also be affected by the performance of the infrared sensor in the imaging system [12-15]. Neural networks offer exceptional solutions for improving images and videos obtained from imperfect meta-optical systems. Previous studies, including our earlier work, have shown respectable performance in enhancing infrared metalenses images using deep learning algorithms such as Cycle-GAN, U-Net, and Pix2Pix, but it does not facilitate high-definition video-rate enhancement for infrared metalenses [16]. In pursuit of high-definition and smooth video enhancement, along with meeting strict quality standards for individual frame images, accurately extracting spatio-temporal information within the video frame sequence is essential [17]. This extraction process aids in precise prediction and interpolation of motion, thereby minimizing artifacts such as jitter or stuttering. 3D convolutional algorithms extend 2D image convolution into the temporal domain by applying sliding convolutions across sequences of video frames to extract temporal features [18-19]. The Ping-Pong loss aims to enhance long-term temporal consistency in video generation through bidirectional loss formulations and self-supervised learning techniques, thereby mitigating the accumulation of artifacts over time [20-21]. RCNN-based video generation networks such as BRCN and RISTN leverage recurrent neural networks (RNNs) to model both temporal dependencies and spatial features [22-23]. These methods operate without requiring explicit alignment of video sequences but may have challenges in accurately computing temporal information. Motion predictors are employed to compensate for predicted motion vectors, ensuring smooth and coherent video outputs [24-25]. Optical flow techniques integrated into video generation models such as Mocycle-GAN, FRVSR, TecoGAN, and VESPCN, are designed to estimate motion between consecutive frames, thereby enhancing the creation of natural and fluid motion in generated videos [26-32].

The performance of the aforementioned methods has demonstrated remarkable results on datasets such as Viper and Flower Video Dataset. However, the extraction of temporal information heavily relies on high-quality datasets. Enhancing the metalens video using these methods proves ineffective due to the considerable impact of chromatic aberration on the imaging signal-to-noise ratio [33-36]. To make the metalens images suitable for video enhancement, we integrated the High-Frequency-Enhancing (HFE) Cycle-GAN with the infrared metalens, resulting in a Neural-Network-Enhanced (NNE) metalens camera. Through high-frequency adversarial training, the generator outputs the reconstructed images that enhance the high-frequency components in the original metalens images by learning the prior knowledge from the corresponding infrared lens images. The adversarial training process encourages the generator to excel at reproducing high-frequency detail, which is critical for restoring the sharp edges, fine details, and intricate patterns. High-frequency enhancement training significantly improves the understanding and feedback of achromatic metalenses images, thereby refining its output results. This results in consistent restoration across successive video frames, even under the challenging low-contrast, low-resolution conditions typical of infrared metalens video, making the NNE metalens camera compatible with commercial infrared cameras.

**The Neural-Network-Enhanced (NNE) Metalens Camera**

The naked infrared metalens is susceptible to chromatic aberration, which results in a significant deviation from the intended target frequency spectrum. The deficiency in low-frequency components contributes to a blurred appearance in the metalens images, which hinders clarity, while the lack of high-frequency components leads to a loss of intricate detail and sharp transitions. Our proposed miniature NNE metalens camera in Figure 1(a) is composed of an infrared metalens, the HFE Cycle-GAN, and the detector. The HFE Cycle-GAN consists of three key modules: Cycle 1, Cycle 2, and High-Frequency Adversarial Learning. Cycle 1 and Cycle 2 are trained cyclically, and the High-Frequency Adversarial Learning module is dedicated to improving the high-frequency recovery in Cycle 1. Due to the aberrations and noise, the frequency spectrum of the metalens image differs greatly from that of the infrared lens image in Figure 1(b), resulting in the original scene being indistinguishable. By establishing the degradation relationship between low-quality naked metalens images and high-quality infrared images, the HFE Cycle-GAN accurately reconstructs the target frequency, thereby correcting the aberrations and reducing noise during metalens imaging.

As shown in Figure 1(c)-(e), the metalens we designed is composed of silicon nanopillars arranged on a square lattice. The input light is converged by the metalens. The required phase of each nanopillar at position (x, y) satisfies the following equation:

$$\varphi(x, y, \lambda) = \frac{2\pi}{\lambda}\left(\sqrt{x^2 + y^2 + f^2} - f\right) \quad (1)$$

where λ=9.5 μm and $f$ =7 mm. The diameter of the metalens is 7 mm. To achieve an entire 2π phase coverage, the height of silicon nanopillars should be comparable to the operating wavelength. In this study, the height of the silicon nanopillars has been set at 5.8 μm. The diameter of a nanopillar varies from 0.5 to 3.5 μm. The metalens is capable of correcting spherical aberration at 9.5 μm. The detailed design and performance of the metalens can be found in Figure S1−S2 of Supporting Information.

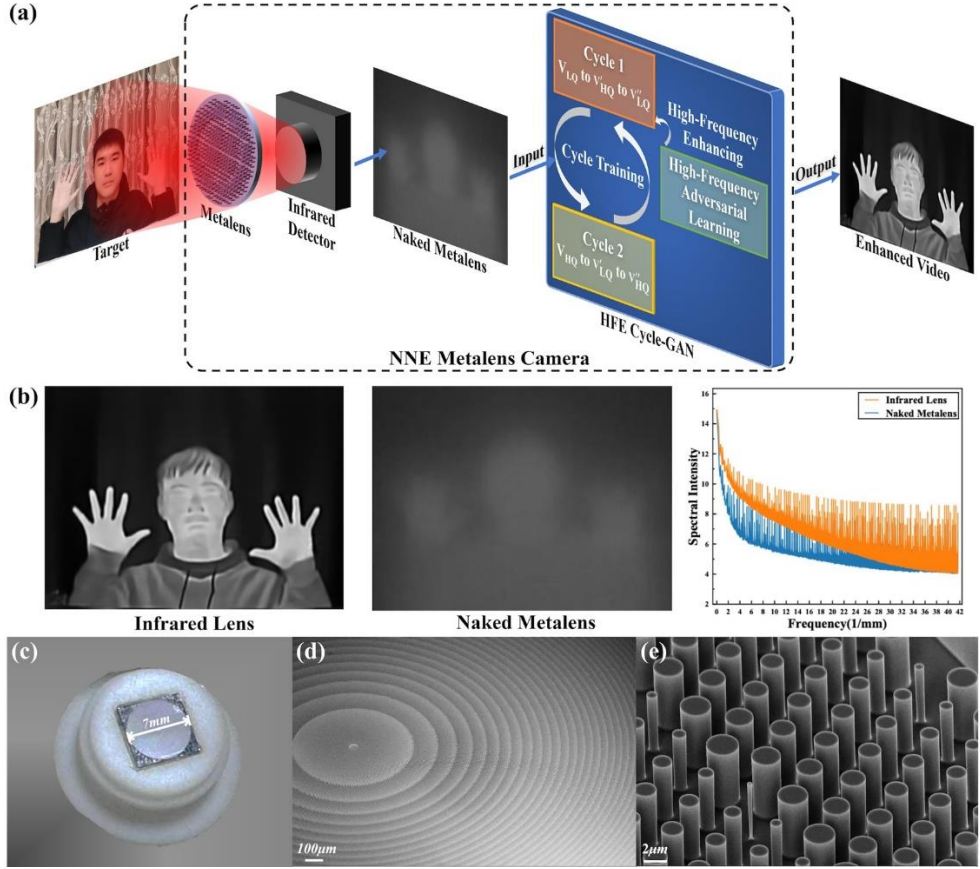

Fig. 1. The Neural-Network-Enhanced (NNE) metalens camera. (a) The NNE metalens camera configuration includes an infrared metalens, the HFE Cycle-GAN for high-frequency enhancement, and an infrared detector. (b) illustrates portrait images captured by commercial infrared cameras and naked metalenses, along with a comparison of their frequency comparison. (c) The infrared metalens attached to a 3D printed component. (d) and (e) show the top views of the infrared metalens, with a scale bar of 2 μm and 100 μm, respectively.

## The High-Frequency-Enhancing (HFE) Cycle-GAN

To perform image-to-image translation, the HFE Cycle-GAN requires a paired dataset of images from two domains. As shown in Figure 2(a), one domain consists of low-quality naked infrared metalens video ($V_{LQ}$) captured by Detector1, while the other domain consists of high-quality infrared video ($V_{HQ}$) captured by Detector2. The HFE Cycle-GAN consists of three modules: Cycle 1, Cycle 2, and High-Frequency Adversarial Learning, shown in Figure 2(b). It includes two generators, the target generator G and auxiliary generator F, and three discriminators: the full-frequency discriminators $D_X$, $D_Y$, and the high-frequency discriminator $D_{HF}$.

In Cycle 1, the $V_{LQ}$ frame is fed into G and undergoes processing through an encoder-decoder structure, depicted in Figure 2(c), comprising four components: Downsampling, Residual Blocks, Upsampling, and Output. The feature of the $V_{LQ}$ frame at layer i ($0 \leqslant i \leqslant 13$) in G is defined as $X_i$, where $X_0$ is the initial $V_{LQ}$ frame. The final output is a high-quality generated video ($V'_{HQ}$) frame. The operations for each layer are described in Equation (2), with the parameters for each operation detailed in Table 1.

$$X_{i+1} = \begin{cases} ReLU(InstanceNorm2D_1(Conv2D_1(X_i))), & 0 \leq i \leq 2 \quad (2.1) \\ X_i + ReLU(InstanceNorm2D_2(Conv2D_2(X_i))), & 3 \leq i \leq 10 \quad (2.2) \\ ReLU(InstanceNorm2D_1(Conv2D_2(DeConv(X_i)))), & 11 \leq i \leq 12 \ (2.3) \\ Tanh(Conv2D_2(X_i)), & i = 13 \quad (2.4) \end{cases} \quad (2)$$

The downsampling contains three layers with the same structure, and the operation of each layer is shown in Equation (2.1). Each layer extracts the features of $X_i$ respectively by convolution operations. Equation (2.2) describes nine Residual Blocks, which aid in conveying features and gradients, thereby deepening the network structure to preserve the shallow features from $V_{LQ}$ frames. Next, the spatial dimensions of the $V_{LQ}$ frames are reconstructed in the Upsampling process defined as Equation (2.3). The output layer defined in Equation (2.4) maps the pixel values of $X_i$ back to the range of the original $V_{LQ}$ frame and outputs $V'_{HQ}$ utilizing the Tanh activation function. F has the same structure as G but operates in the opposite direction. It takes the translated $V'_{HQ}$ frame and converts it back to the $V'_{LQ}$ frame, similar to the original $V_{LQ}$ frame. This cyclic generative structure enables support for unpaired datasets, ensuring effective image translation and reconstruction.

**Table 1. The neural network layers comprise the generators and discriminators with the corresponding parameters for each layer.**

| Model | Layer | Parameter | Numbers |
|---|---|---|---|
| Generators | $Conv2D_1$ | 7 × 7, stride = 2, padding = 1 | 2 |
| | $Conv2D_2$ | 3 × 3, stride = 2, padding = 1 | 13 |
| | DeConv | scale_factor = 2 | 2 |
| | $InstanceNorm2D_1$ | features = 64 | 15 |
| | ReLU | - | 15 |
| | Tanh | - | 1 |
| Discriminators | $Conv2D_3$ | 4 × 4, stride = 2, padding = 1 | 4 |
| | $InstanceNorm2D_2$ | features = 128/256/512 | 1 |
| | LeakyReLU | negative_slope = 0.2 | 4 |

In Cycle 1, $D_Y$ is paired with G and acts as a full frequency discriminator. $D_Y$ distinguishes between the $V_{HQ}$ frame and the $V'_{HQ}$ frame by assessing features such as edges, textures, and patterns using a series of convolutional layers defined as:

$$Y_{i+1} = LeakyReLU\left(InstanceNorm2D_2(Conv2D_3(Y_i))\right), \quad 0 \leq i \leq 3 \quad (3)$$

where $Y_i$ represents the feature of the $V_{HQ}$ or $V'_{HQ}$ frame at the layer i ($0 \leqslant i \leqslant 3$), and $Y_0$ is the initial input frame. The final $Conv2D_3$ layer generates a 70×70 matrix by identifying subtle differences between the $V_{HQ}$ frame and $V'_{HQ}$ frame. Each element of this matrix represents a probability score for each patch, indicating the likelihood that the local region within the patch is real. Each patch represents a small region of the input frame. In Cycle 2, F and Dx are trained similarly to Cycle 1. F takes the $V_{HQ}$ frame as input and outputs the low-quality infrared metalenses video ($V'_{LQ}$) frame. On the other hand, G generates $V'_{HQ}$ frames, which are reconstructed frames corresponding to $V_{HQ}$, thus establishing a cyclic structure similar to Cycle 1 but operating in the opposite direction.

Due to the chromatic aberration and noise, the metalens images almost have no detail. To further compensate for the loss of detail in metalens imaging, we extracted the high-frequency components using wavelet transform and conducted additional high-frequency adversarial learning on them. As depicted in Figure 2(b), the high-frequency band of the $V'_{HQ}$ is decomposed using a two-dimensional Discrete Wavelet Transform (DWT) employing the Haar wavelet function:

$$(LL, HL, LH, HH) = DWT_{Haar}(V'_{HQ}) \qquad (4)$$

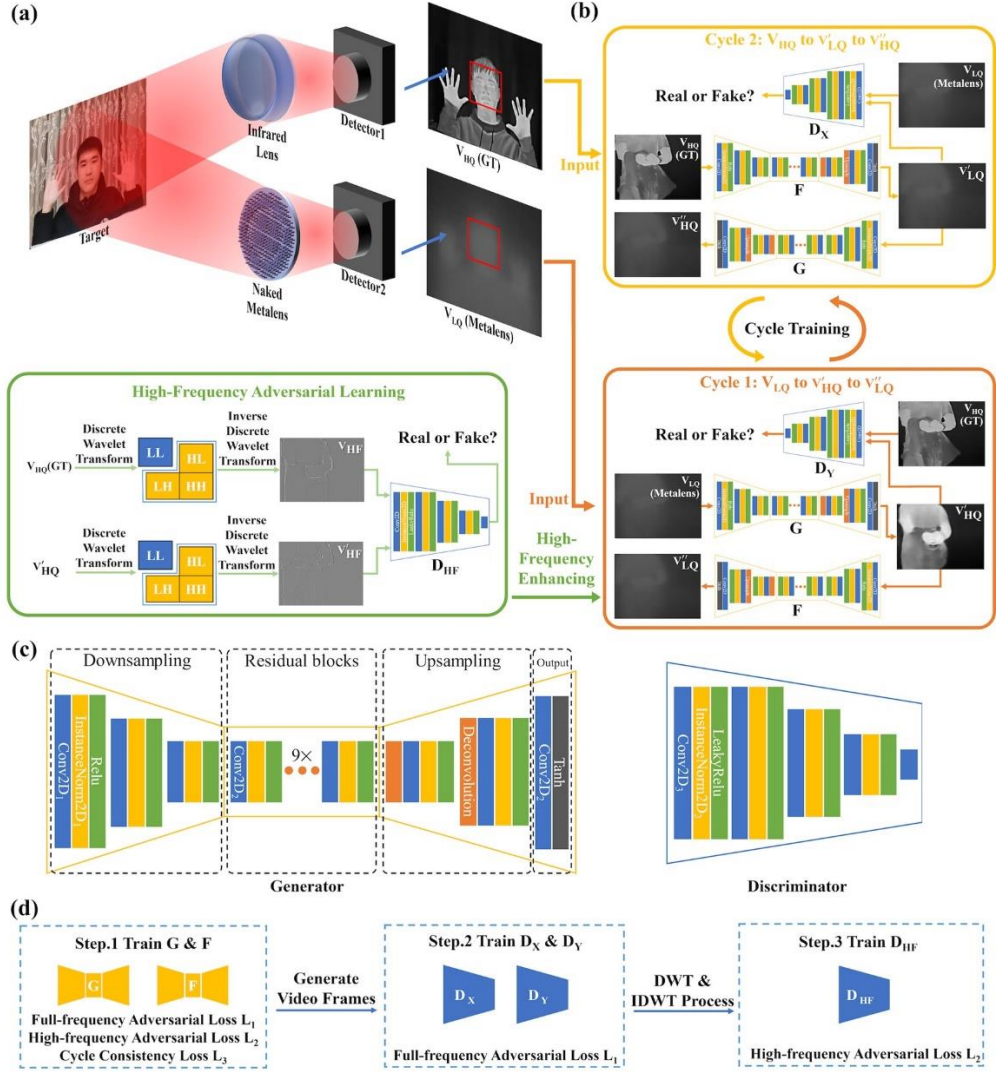

Fig. 2. Structure and training process of the High-Frequency-Enhancing (HFE) Cycle-GAN. (a) The process of acquiring pairs of high-quality ($V_{HQ}$) and low-quality ($V_{LQ}$) videos. $V_{HQ}$ videos are captured using the commercial infrared camera, while $V_{LQ}$ videos are captured using the infrared metalens camera. (b) The HFE Cycle-GAN consists of Cycle 1, Cycle 2, and the High-Frequency Adversarial Learning module. Cycle 1 comprises G, F, and $D_Y$ which perform the conversion from $V_{HQ}$ to $V'_{LQ}$ and back to $V''_{HQ}$. Cycle 2 comprises G, F, and $D_X$ facilitating the conversion from $V_{LQ}$ to $V'_{HQ}$ and then back to $V''_{LQ}$. The high-frequency discriminator $D_{HF}$ in the High-Frequency Adversarial Learning module assesses the high-frequency components separated from the video frame, excluding the LL band. (c) The structure of the generator and discriminator. The generator consists of a downsampling, residual block, upsampling, and output layer. The discriminator utilizes a 5-layer convolutional structure. (d) Each training epoch involves three main steps: (1) generating $V_{HQ}$ and $V_{LQ}$ video transformations, (2) updating discriminators to evaluate full and high-frequency quality, and (3) adjusting generators to minimize adversarial loss, cycle-consistency loss, and high-frequency loss.

This decomposition yields four frequency bands: HH (Diagonal Detail Coefficients), HL (Horizontal Detail Coefficients), LH (Vertical Detail Coefficients), and LL (Approximation Coefficients), where the HH, HL, and LH make up the high-frequency band. This high-frequency band is converted back to the spatial-domain video ($V'_{HF}$) frame using the two-dimensional Inverse Discrete Wavelet Transform (IDWT):

$$V'_{HF} = IDWT(ZERO, HL, LH, HH) \tag{5}$$

ZERO refers to an all-zero matrix that is used to replace the original LL band to preserve the high-frequency component of the video frame. The $V'_{HF}$ selectively preserves the high-frequency components of the $V_{HQ}$ frame, allowing for targeted enhancement during further training processes focused on the high-frequency components.

A similar process is applied to derive the high-frequency band video ($V_{HF}$) frame from the corresponding $V_{HQ}$ frame. Both the $V'_{HF}$ frame and $V_{HF}$ frame are then fed into $D_{HF}$ for adversarial learning alongside G in Cycle 1.

In High-Frequency Adversarial Learning, we proposed the high-frequency adversarial loss $L_1$ for $D_{HF}$, specifically designed to enhance the generative capabilities of the generator G within the high-frequency bands, Equation (6):

$$L_1 = \sum_t \log\left(D_{high}(V_{HF})\right) + \sum_t \log\left(1 - D_{high}(V'_{HF})\right) \tag{6}$$

It establishes a zero-sum game where the generator G aims to maximize its probability of effectively deceiving $D_{HF}$, while $D_{HF}$ strives to minimize the likelihood of being deceived. With fixed parameters for $D_{HF}$, G generates video frames designed to prompt $D_{HF}$'s output to approximate 1, thereby guiding the $V'_{HF}$ frame to converge towards the original $V_{HF}$ frame. Simultaneously, with the parameters of G fixed, $D_{HF}$ is trained to output values close to 1 for the $V_{HF}$ frame and 0 for the $V'_{HF}$ frame. This dual training approach is essential for achieving high-frequency adversarial learning in G. Through adversarial training, $D_{HF}$ provides detailed feedback to G regarding the fidelity and accuracy of high-frequency features in the enhanced video frames. This feedback loop helps G refine its outputs, resulting in smoother transitions and improved coherence between frames.

$L_2$ is the full-frequency adversarial loss function used to train G, F, and $D_x$, $D_y$. The video frames generated by the generators and input to the discriminators encompass the full spectrum of frequencies. By combining the $L_1$ and $L_2$ loss functions, G achieves the capability to generate higher-quality infrared video frames with enhanced detail.

$$L_2 = \sum_t \log\left(D_Y(V_{HQ})\right) + \sum_t \log\left(1 - D_Y(V'_{HQ})\right) + \sum_t \log\left(D_X(V_{LQ})\right) + \sum_t \log\left(1 - D_X(V'_{LQ})\right) \tag{7}$$

The cycle consistency loss $L_3$ maintains the content consistency of video frames during domain translations by imposing bidirectional constraints. It achieves this by calculating the Manhattan Distance between $V_{LQ}$ frames, $V_{HQ}$ frames, and their respective reconstructed counterparts $V''_{LQ}$ frames and $V''_{HQ}$ frames throughout the translation process.

$$L_3 = \sum_t (V_{LQ} - V''_{LQ}) + \sum_t (V_{HQ} - V''_{HQ}) \tag{8}$$

The loss function L is defined as Equation (9):

$$L = \omega_1 L_1 + \omega_2 L_2 + \omega_3 L_3 \tag{9}$$

We implemented our network using the PyTorch framework. The training process for each epoch involves three main steps, as illustrated in Figure 2(c). The parameters of discriminators $D_X$, $D_Y$, and $D_{HF}$ parameters are fixed at first. G and F generate the $V'_{HQ}$ frame and $V'_{LQ}$ frame, respectively, using the $V_{LQ}$ frame and $V_{HQ}$ frame as inputs. The parameters of G and F are then optimized based on $L_1$ and $L_2$. $L_3$ facilitates the reconstruction of $V''_{HQ}$ frames and $V''_{LQ}$ frames using G and F. The parameters of $D_X$ and $D_Y$ are subsequently updated using $L_2$, which is designed to enhance their discriminative capabilities while keeping the parameters of G and F fixed. The parameters of $D_{HF}$ are optimized using $V_{HF}$ frames and $V'_{HF}$ frames and guided by $L_1$. Following optimization, trade-off parameters are set to $\omega1=5$, $\omega2=10$, and $\omega3=5$ and the optimization process of the hyperparameters can be found in the Supporting Information Figure S3. Through high-frequency adversarial learning, the $V'_{HF}$ frames generated by G align more closely with the $V_{HF}$ frames. This means that the HFE Cycle-GAN, by emphasizing the high-frequency spectrum, provides a deep understanding of the degradation of metalens images. The high-frequency adversarial learning mitigates the lack of sharpness and complexity evident in the captured metalens video.

Ensuring robust training of the HFE Cycle-GAN requires extensive prior knowledge and a variety of image degradation types throughout the end-to-end learning process. Therefore, the dataset we created includes a variety of scenarios such as people, buildings, road traffic, and other diverse environments to ensure a broad representation of real-world conditions. The cameras record 24 frames per second. Each video frame is standardized to a grayscale image of dimensions $1\times256\times192$ and processed in batches of size 10. More information about the acquisition environment of the dataset as well as the pre-processing process and loss curve can be found in the Supporting Information Figure S4. It contains a total of 19,715 pairs of video frames extracted from thirty recorded clips. The dataset is split into a 9:1 train-test ratio. By comparing the frequency components of $V_{LQ}$ and $V_{HQ}$ frames, the HFE Cycle-GAN can understand the frequency degradation model present in metalens imaging, and effectively target and reconstruct the distorted frequency spectra.

**Resolution Calibration**

To assess the high-frequency enhancement capability of the NNE metalens camera, we quantified the resolution by using a calibration board. This board is fabricated with ceramic material and features slits of varying widths ranging from 2 mm to 6 mm, as shown in Figure 3(a). The slits are indistinguishable from the naked metalens, whereas the NNE metalens camera can resolve slits with a minimum width of 2 mm at a distance of 1 m, with an angular resolution of 0.002°. The spatial resolution of the NNE metalens camera is on par with that of the commercial infrared lens camera (GT). To further validate the effectiveness of the NNE metalens camera in object restoration, we demonstrate the reconstructed human faces, upper limbs, and outdoor scenes in Figure 3(b). The reconstructed facial features show remarkable clarity and definition, capturing even intricate details such as the distinct texture of hair. In addition, the texture of elements such as trees and buildings are presented with exceptional clarity, similar to images captured by an infrared camera.

To demonstrate the effectiveness of the HFE Cycle-GAN in high-frequency enhancement, we employed Cycle-GAN as a comparative network for ablation experiments. This allows for a direct comparison to HFE Cycle-GAN, as both networks share the same generator and discriminator structure, with the primary distinction being the absence of the high-frequency adversarial learning module in Cycle-GAN. In Figure 3(c), we calculated the average spectral intensity of 200 randomly selected frames. In the low frequencies below 10 $mm^{-1}$, both the HFE Cycle-GAN and Cycle-GAN reconstructed frames exhibit comparable average frequency intensity, indicating that both methods can accurately restore the general contours of the metalens images. However, in the high-frequency band, especially in the part with frequencies higher than 15 $mm^{-1}$, the HFE Cycle-GAN outperforms Cycle-GAN by showing a closer resemblance to the frequency of the ground truth. This suggests that our HFE Cycle-GAN has a deeper understanding of the imaging characteristics of the metalens and thus effectively reduces the frequency difference between the $V_{LQ}$ frames and the $V_{HQ}$ frames. The inclusion of the high-frequency adversarial learning module in the HFE Cycle-GAN notably enhances the fine details of the $V_{LQ}$ frames, aligning with the findings of the resolution calibration mentioned above.

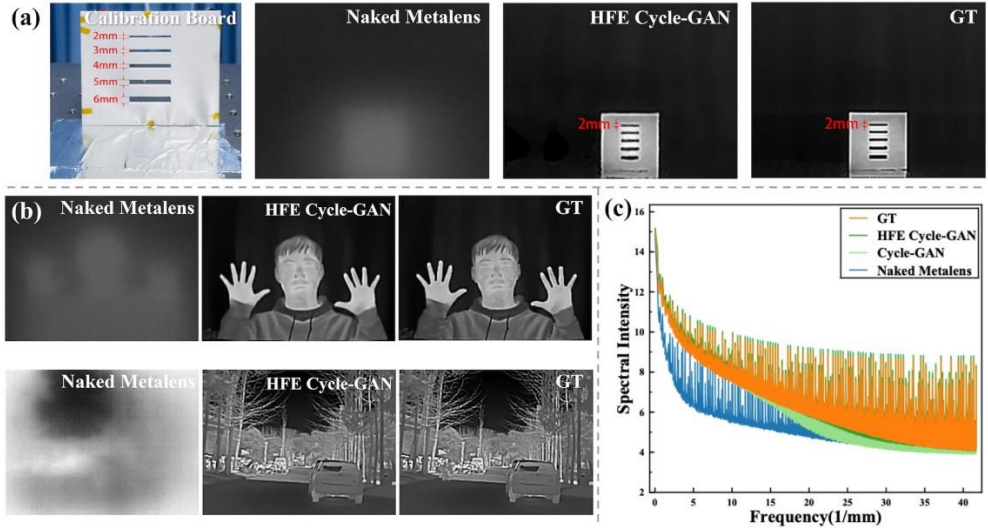

Fig. 3. Resolution quantification of the HFE Cycle-GAN. (a) depicts the calibration board image captured by the visible light camera, naked metalenses, and commercial infrared camera, along with the image reconstructed by the HFE Cycle-GAN. (b) depicts the reconstructed the metalenses images of human faces, upper limbs, and outdoor scenes. (c) depicts the average frequency intensity of 200 video frames. The darker green curve (the metalens image reconstructed by the HFE Cycle-GAN) overlaps closely with the orange curve (GT).

**Image Qualification**

The accuracy and fidelity of the translated video frames are assessed using traditional metrics such as Peak Signal to Noise Ratio (PSNR), Structure Similarity Index Measure (SSIM), and Fréchet Inception Distance (FID), with the aim of improving the assessment of perceptual quality for GAN-based networks by quantifying the disparity between feature-level distributions. The reconstructed video frames from HFE Cycle-GAN are compared with those from three state-of-the-art networks: Cycle-GAN, Mocycle-GAN, and RISTN. To ensure a fair and direct assessment, both Cycle-GAN and Mocycle-GAN employ the same generator and discriminator architecture as our HFE Cycle-GAN. In Figure 4-6, by including fifteen consecutive frames from each of the three test videos, we can assess the performance of different networks in different scenarios. It is noteworthy that our method performs consistently well across all scenarios. The frames from Video 1 shown in Figure 4 illustrate that the HFE Cycle-GAN excels at enhancing intricate, high-frequency details such as the fine arm and sleeve details of clothing, as well as complex elements such as folds and watches. In contrast, Cycle-GAN, Mocycle-GAN, and RISTN restore blurred outlines and struggle to capture these high-frequency details, resulting in a less detailed restoration. When observing the frames of Video 2 and Video 3 in Figure 5 and Figure 6, the HFE Cycle-GAN shows superior performance in reproducing intricate details, especially in structures such as buildings, trees, and roads. Conversely, the Cycle-GAN, Mocycle-GAN, and RISTN outputs are less accurate, with barely discernible building outlines and an overall lack of detail. This enhancement capability of the HFE Cycle-GAN is attributed to the refined handling of high-frequency details resulting from the enhanced high-frequency training process of our method.

The radar charts in Figure 4-6 demonstrate the average results of PSNR, SSIM, and FID of the video frames from three test videos. The size of the polygon in the radar charts directly indicates the quality of the video frames, with larger areas signifying better performance. The HFE Cycle-GAN consistently demonstrates larger areas in the radar charts, clearly affirming its superior capability in enhancing video quality across all metrics. For example, in Figure 4(b), the HFE Cycle-GAN improves the PSNR of the $V_{LQ}$ frames by 18.26%. Compared to the Cycle-GAN, the HFE Cycle-GAN shows an 8.76% improvement in PSNR and a 15.79% improvement in SSIM as shown in Figure 4(c). In contrast, the Mocycle-GAN and RISTN only show a modest improvement of 1.08% and 3.89% in PSNR, 10.53% and 3.39% in SSIM. Different from other comparison networks, the PSNR and SSIM results for the HFE Cycle-GAN's reconstructed videos across different scenes remain relatively stable, highlighting the network's robust generalization capability. The lack of a one-to-one pixel match between the ground truth infrared image and the enhanced image makes it difficult to accurately assess image quality using PSNR and SSIM. In addition, PSNR and SSIM may not fully capture the human perception of video image quality. To overcome these limitations, we use FID to improve the assessment of perceptual quality for GAN-based networks by quantifying the disparity between two feature-level distributions. In Figure 5, the FID of Video 2 indicates that Mocycle-GAN and RISTN achieve a 10.59% and 17.65% improvement over Cycle-GAN. In contrast, HFE Cycle-GAN demonstrates an even more remarkable 50.59% improvement. In comparison to $V_{LQ}$ frames, HFE Cycle-GAN shows substantial enhancements in recovering video frames, particularly at the depth feature level. These findings indicate that the HFE Cycle-GAN is superior to existing video translation networks in terms of capturing and reproducing detailed features, thereby significantly enhancing image quality. More test videos are available in Supporting Information Figure S3 and Supporting Information Video 1-5.

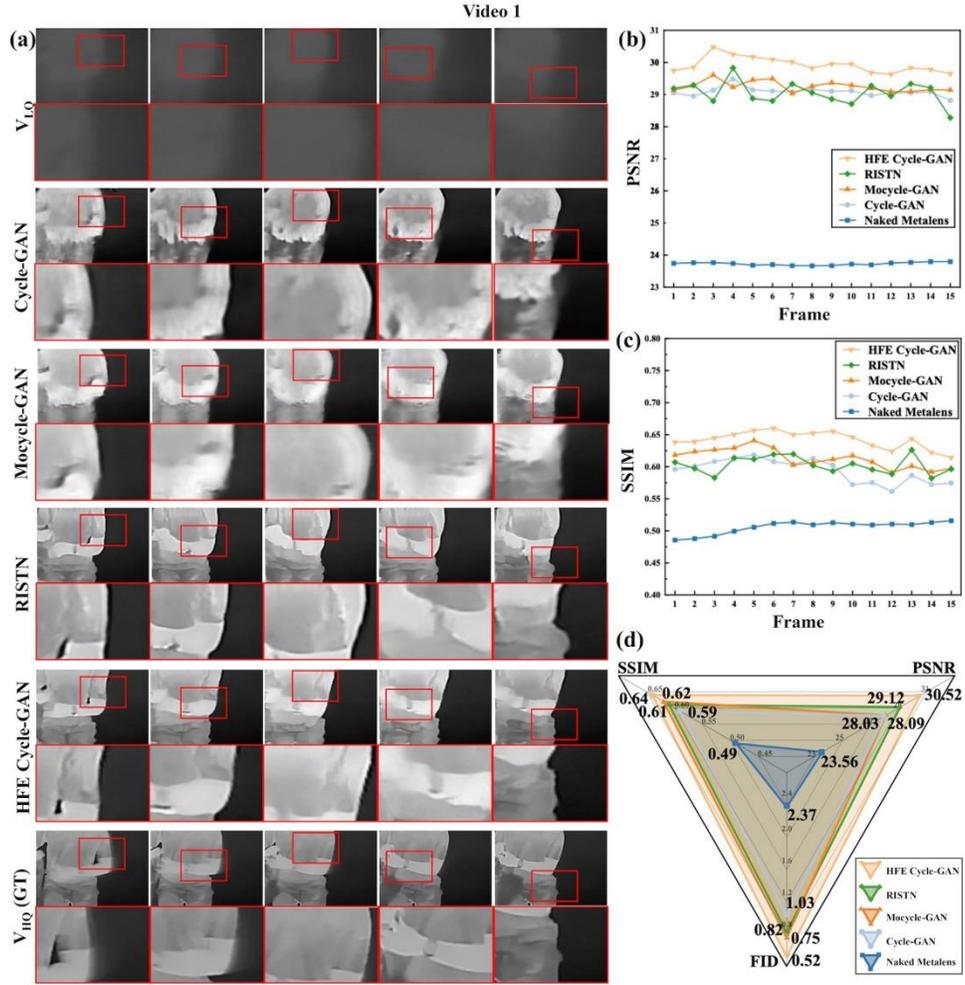

Fig. 4. The evaluation of Video 1. (a) illustrates five consecutive frames from Video 1. The Video reconstructed by the HFE Cycle-GAN captures the fine details in clothing and accessories. (b) and (c) show PSNR and SSIM across 15 consecutive frames, and (d) presents the average PSNR and SSIM over 100 frames, demonstrating that the HFE Cycle-GAN outperforms the comparison networks in both metrics.

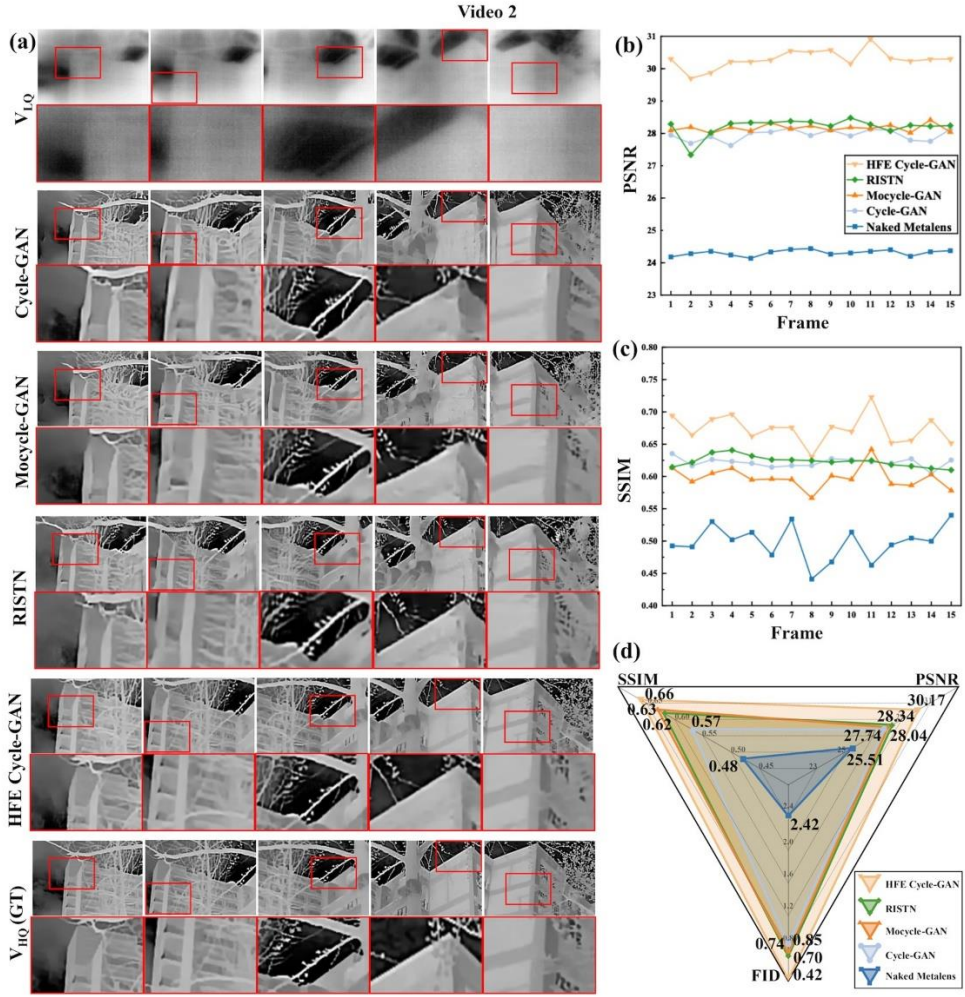

Fig. 5. The evaluation of Video 2. (a) presents five frames from Video 2, which showcases the reconstruction results of an outdoor scene, including building edges and foliage. (b) and (c) display PSNR and SSIM across 15 frames, and (d) shows the average metrics over 100 frames. In terms of PSNR, the HFE surpasses the less effective RINST by an 6.46%.

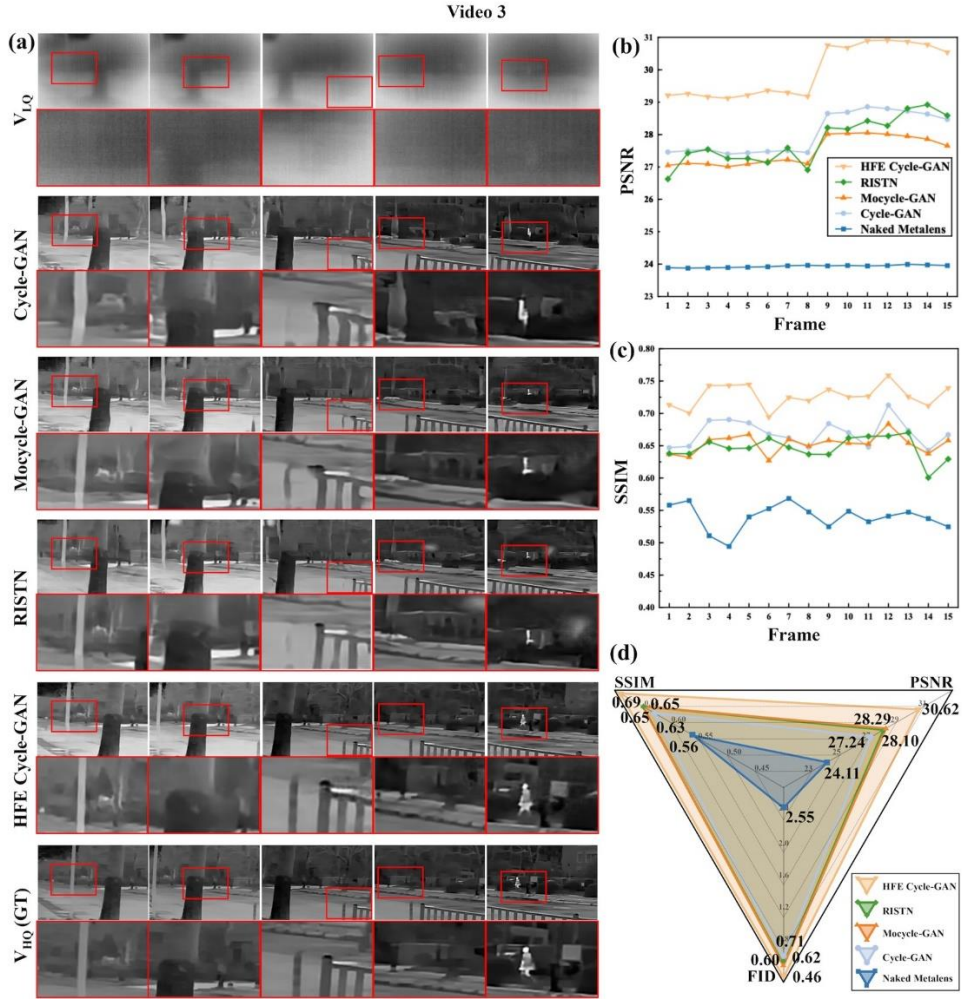

Fig. 6. The evaluation of Video 3. (a) presents five frames from Video 3, which showcases the reconstruction results of roads and trees. (b) and (c) display PSNR and SSIM across 15 frames, and (d) shows the average metrics over 100 frames.

**Video Smoothness**

High image quality is the basis for smooth video output. Metrics such as SSIM, PSNR, and FID concentrate on image-level assessment, neglecting the human perception of video smoothness. We presented a video smoothness evaluation method based on optical flow. This calculates the speed of pixel movement and displacement between consecutive frames, generating a vector field with direction and length by extracting the optical flow motion vector from the enhanced video, which can provide insights into the smoothness and motion dynamics of the video and compares it to the optical flow motion vector of GT, calculating the error using Euclidean distance (End Point Error or EPE). Equation (10) defines the EPE as the Euclidean distance between the horizontal (u, u') and vertical (v, v') components of the optical flow fields. Lower EPE values indicate smoother video results.

$$EPE = \sqrt{(u - u')^2 + (v - v')^2} \qquad (10)$$

Figure 7 presents the results of optical flow estimation for different enhancement networks. In Figure 7(a), white in the color-coded flow field indicates minimal movement. If the video is smooth, such as the GT video, the color-coded flow field of the background should be white to indicate that the background is stationary. Similarly, the HFE Cycle-GAN yields comparable results, with the color-coded flow field of the background being predominantly white. In contrast, the Cycle-GAN and Mocycle-GAN exhibit darker colors of the background in the color-coded flow field, indicating instability in restoring the background across consecutive video frames. Also, when compared to other networks, the HFE Cycle-GAN demonstrates a higher similarity to the ground truth in predicting the motion of the body in the color-coded flow field. In Figure 7(b), we calculate the EPE for consecutive fifty frames in each test video. The HFE Cycle-GAN consistently delivers lower EPE results when compared to both Cycle-GAN and Mocycle-GAN. Furthermore, we calculate the average EPE for the three distinct videos. As depicted in Table 2, the HFE Cycle-GAN demonstrates a substantial 53.04% improvement over the Mocycle-GAN network. This improvement underscores the advantageous accuracy and stability in luminance translation that our method offers, resulting in notably smoother translation outcomes.

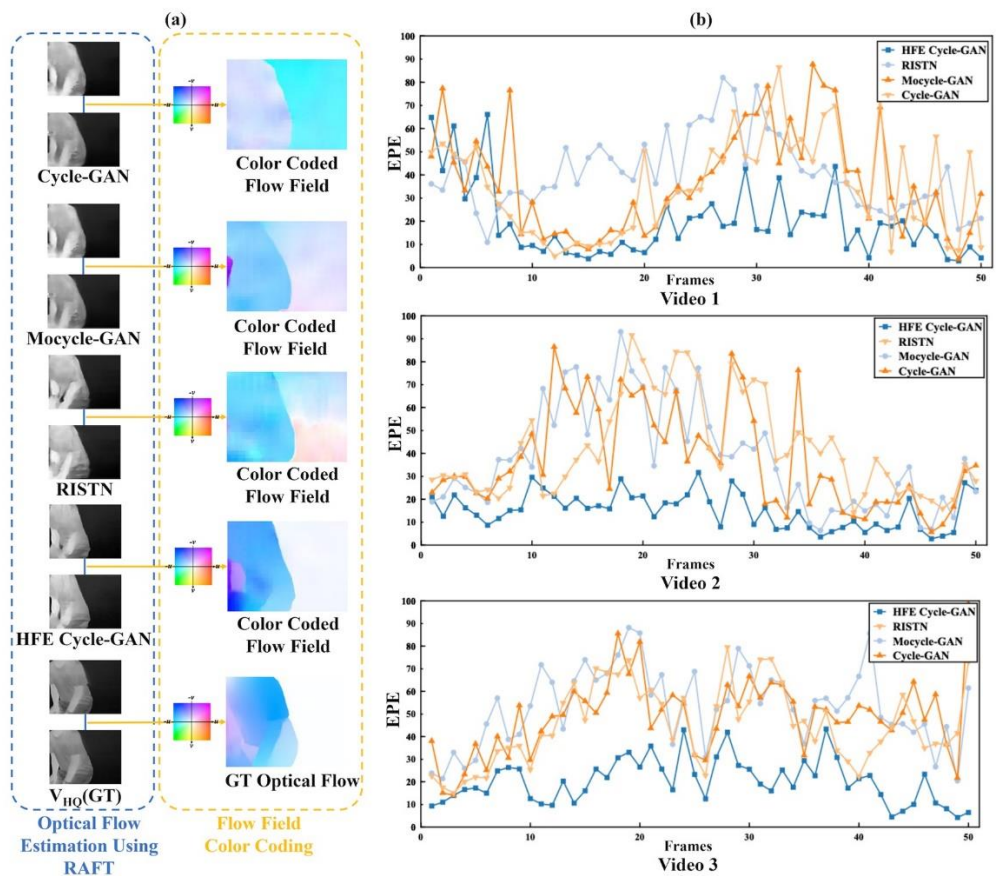

Fig. 7. Evaluation of video smoothness using optical flow. (a) demonstrates the usage of RAFT (Recurrent All-pairs Field Transforms) for optical flow estimation in metalens video sequences, along with the generation of a color-coded flow field map. The color-coded flow field map illustrates motion vectors for each pixel, with different colors representing both the direction and magnitude of motion relative to the center point. (b) depicts the results of EPE computation for three test videos. Each chart shows the results of fifty consecutive video frames of one test video. The dark blue curve (EPE of the HFE Cycle-GAN) remains below 50 on the EPE values for most frames, displaying smoother variations. In contrast, the other curves exhibit larger fluctuations in their values.

Table 2. The average EPE reconstructed by different networks

| Model | Video 1 | Video 2 | Video 3 | Video 4 | Video 5 |
|---|---|---|---|---|---|
| Cycle-GAN | 27.24 | 40.11 | 47.62 | 23.94 | 34.66 |
| Mocycle-GAN | 26.13 | 36.75 | 46.55 | 23.52 | 31.52 |
| RISTN | 26.97 | 35.24 | 47.10 | 22.90 | 33.89 |
| HFE Cycle-GAN | 14.24 | 19.21 | 21.03 | 12.58 | 26.57 |

In addition to the objective evaluation through optical flow analysis, we conducted a subjective evaluation to comprehensively assess the performance of the enhanced video based on human perception. In these experiments, 152 participants were presented with the enhancement results of three networks simultaneously. They were then asked to select the best video based on their sensory experience. To ensure impartiality, the order of the test videos was randomized, and the participants were unaware of the enhancement method used for each video. The aggregated results support our approach, with over 90% of participants agreeing that our results were more perceptually realistic and smoother, as detailed in Table 3. These results are consistent with the results of the optical flow-based evaluation, demonstrating that the HFE Cycle-GAN has superior performance in infrared metalens video enhancement. The NNE metalens camera, empowered by HFE Cycle-GAN, achieves real-time high-definition video output at a frame rate of 8ms.

Table 3: Subjective evaluation of the reconstructed video based on human perception, where higher scores indicate better perceived quality

| Method | Video 1 | Video 2 | Video 3 | Video 4 | Video 5 |
|---|---|---|---|---|---|
| Cycle-GAN | 1.95% | 1.95% | 0.65% | 1.30% | 0.13% |
| Mocycle-GAN | 3.25% | 1.95% | 1.30% | 1.95% | 0.65% |
| HFE Cycle-GAN | 94.81% | 96.10% | 98.05% | 96.75% | 93.40% |

**Conclusion**

Our work presents a compact NNE metalens camera consisting of infrared metalens, a CMOS image sensor, and the HFE Cycle-GAN. Based on the prior knowledge gained from infrared lens images, the HFE Cycle-GAN understands the frequency degradation that metalens undergo. This understanding facilitates the mitigation of losses due to aberrations and noise, resulting in real-time high-resolution imaging of the NNE metalens. The HFE Cycle-GAN exhibits strong robustness in various indoor and outdoor scenes. It simplifies the design and manufacturing processes of metalenses by alleviating the stringent requirements usually associated with creating an ideal metalens, compensating for imperfections through end-to-end image restoration. This simplification encourages wider adoption of metalens in various applications, making it more accessible and practical for developers and manufacturers. Furthermore, the application of the HFE Cycle-GAN is not limited to metalens, as it provides a universal solution for video enhancement in faulty optical systems, as long as the dataset is diverse enough for the network to capture the frequency degradation of the optical system. In addition, the video evaluation method proposed in our work enables a quantitative estimation of video enhancement quality, with a special focus on smoothness, by bridging the gap between objective quantitative metrics and subjective human perception.

In theory, both noise and the optical transfer function constrain imaging performance. High noise can introduce high-frequency components that may inadvertently be restored by the HFE Cycle-GAN, leading to artifacts. However, our current study has not quantitatively analyzed the limits of the signal-to-noise ratio of images that could be restored by the HFE Cycle-GAN. While there are algorithms tailored for noise reduction, they often smooth the image, contradicting the novelty of our approach in this paper. Thus, integrating our method with denoising algorithms to improve overall performance warrants further investigation.

**Supporting Information Available:** Camera setup and specifications; structure and fabrication details of the metalens, including phase and transmission ratio profiles; optimization process of the hyperparameters; acquisition environment of the dataset; pre-processing process of the dataset; loss curve of the training process; evaluation results of Videos 4 and 5; and visualizations of Videos 1-5. This material is available free of charge via the Internet at http://pubs.acs.org

**Acknowledgments:** Project supported by Basic Public Welfare Research Program of Zhejiang Province (No. LDT23F05013F05), National Key Research and Development Program of China (2023YFF0613000), National Natural Science Foundation of China (No.62222511), Natural Science Foundation of Zhejiang Province China (LR22F050006) and STI 2030–Major Projects (2021ZD0200401).

**Disclosures:** The authors declare no conflicts of interest.

**Data availability:** Data underlying the results presented in this paper are not publicly available at this time but may be obtained from the authors upon reasonable request.

# Neural-Network-Enhanced Metalens Camera for High-Definition, Dynamic Imaging in the Long-Wave Infrared Spectrum

JING-YANG WEI,[1,†] HAO HUANG,[1,†] XIN ZHANG,[1] DE-MAO YE,[1] YI LI,[1] LE WANG,[1] YAO-GUANG MA,[2,*] AND YANG-HUI LI [1,*]

[1]College of Optical and Electronic Technology, China Jiliang University, Hangzhou 310018, China
[2]State Key Laboratory of Extreme Photonics and Instrumentation, College of Optical Science and Engineering; Intelligent Optics and Photonics Research Center, Jiaxing Research Institute; ZJU–Hangzhou Global Scientific and Technological Innovation Center; International Research, Center for Advanced Photonics, Hangzhou, Zhejiang 310013, China
†These authors contributed equally to this work.
*Correspondence should be addressed to Yao-guang Ma (mayaoguang@zju.edu.cn) and Yang-hui Li (lyh@cjlu.edu.cn).

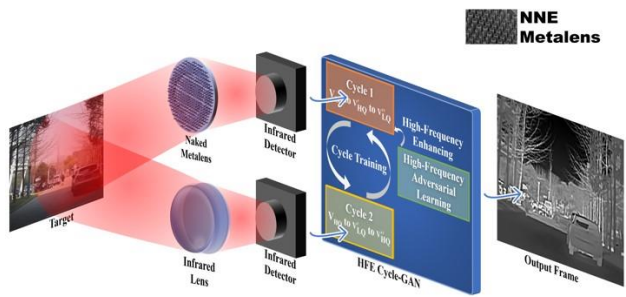

# Compact dynamic imaging system for infrared metalens driven by High-Frequency-Enhancing Cycle-GAN network. Supporting Information: Neural-Network-Enhanced Metalens Camera for High-Definition, Dynamic Imaging in the Long-Wave Infrared Spectrum

JING-YANG WEI,[1,†] HAO HUANG,[1,†] XIN ZHANG,[1] DE-MAO YE,[1] YI LI,[1] LE WANG,[1] YAO-GUANG MA,[2,*] AND YANG-HUI LI [1,*]

[1]*College of Optical and Electronic Technology, China Jiliang University, Hangzhou 310018, China*
[2]*State Key Laboratory of Extreme Photonics and Instrumentation, College of Optical Science and Engineering; Intelligent Optics and Photonics Research Center, Jiaxing Research Institute; ZJU–Hangzhou Global Scientific and Technological Innovation Center; International Research, Center for Advanced Photonics, Hangzhou, Zhejiang 310013, China*
[†]*These authors contributed equally to this work.*
*\*Correspondence should be addressed to Yao-guang Ma (mayaoguang@zju.edu.cn) and Yang-hui Li (lyh@cjlu.edu.cn).*

**Number of pages: 5.**

**Number of figures: 5.**

**Camera Setup**

The $V_{LQ}$ and $V_{HQ}$ frames are captured simultaneously. The $V_{LQ}$ frames are captured by the infrared detector equipped with the metalens, while the $V_{HQ}$ frames are captured by the Dahua DH-TPC-BF2221-B7F8 infrared camera shown in Figure S1. The infrared detector has a resolution of 256×192 pixels, with a pixel size of 12µm. The metalens used in this setup has a focal length of 7 mm and a diameter of 7 mm, operating within a spectral range of 8–14µm. Similarly, the infrared lens of the Dahua DH-TPC-BF2221-B7F8 infrared camera has identical specifications to the metalens.

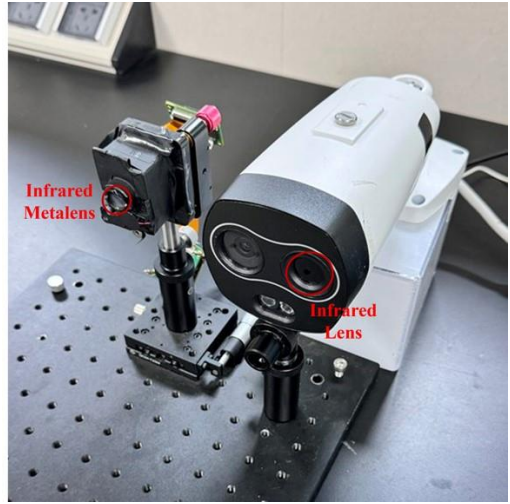

Fig. S1. Infrared dataset acquisition equipment.

**Structure and Fabrication of the Metalens**

The material of the metalens is silicon. The structure of the unit cell of the metalens is as follows: the periodicity (P) is 4 µm, the height (H) is 5.8 µm, and the diameter ranges (D) from 0.5 µm to 3.5 µm. The optimization results of the unit structure satisfy the requirement of $2\pi$ phase coverage, achieving an average transmittance of approximately 69.3%. The specific phase and transmission can be found in Figure S2. We utilized a silicon wafer substrate with a thickness of 700µm for the fabrication of the metalens. The patterns are transferred onto the photoresist layer through laser direct writing (LDW, HEIDELBERG DWL66+). To ensure etching selectivity, we employed a lift-off process. A chromium (Cr) layer measuring 50 nm in thickness is deposited onto the surface of the photoresist, and subsequently, the photoresist layer is eliminated using a stripper solution to successfully transfer the pattern onto the Cr layer. For transferring the pattern onto the Si layer, we employed inductively coupled plasma etching (ICP) with SF6 and C4F8 gases. Finally, by employing a wet etchant, we removed any remaining Cr to obtain our desired samples.

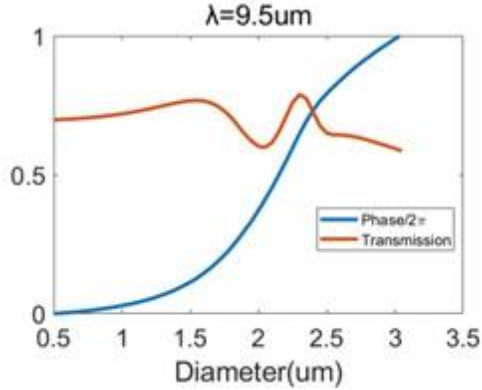

Fig. S2. Phase and transmission ratio profiles of the unitary structure.

**Hyperparameter Optimization**

During the training of the network, we started with baseline values from our previous work: $\omega_2 = 10$ and $\omega_3 = 5$, which are standard for Cycle-GAN. For the high-frequency adversarial loss parameter $\omega_1$, we initially set it to 1 and gradually adjusted it based on the network's performance. After each adjustment, we evaluated the network's ability to restore high-frequency details and the overall quality of the restored images. As shown in Figure S3, when $\omega_1$ is set below 5, the improvement over the original Cycle-GAN is minimal, indicating that higher weight is needed to prioritize high-frequency detail restoration. However, when $\omega_1$ exceeds 5, the network exhibits difficulties in convergence, leading to instability during training. After extensive testing, it demonstrates that $\omega_1 = 5$ offers the best balance between stable convergence and high-frequency detail restoration, resulting in optimal restoration performance.

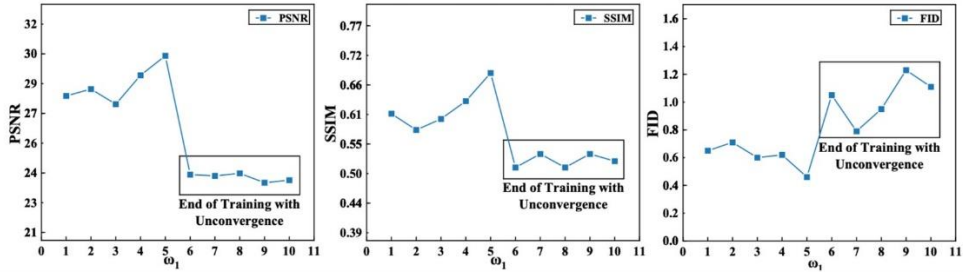

Fig. S3. The effect of different values of $\omega_1$ on model performance. The model achieves its best performance in PSNR, SSIM, and FID metrics when $\omega_1$ is set to 5. For $\omega_1$ values greater than 5, training becomes challenging due to convergence difficulties.

**Dataset and Loss Curves**

To guarantee that the model can function effectively in a wide range of real-world situations, our dataset encompasses a diverse array of indoor and outdoor scenes. The indoor scenes are filmed at room temperature, while the outdoor scenes are captured in varying temperatures and lighting conditions, showcasing complex environments such as buildings, streets, vehicles, and trees. This diversity in settings is essential for capturing a broad range of contrast and noise levels, enhancing the overall variety of the dataset. All video sequences are initially recorded at a resolution of 256×192 pixels per frame, with a frame rate of 24 frames per second. Subsequently, the input videos are resized to 256×256 pixels. To introduce variability in the image regions processed by the network, we applied random cropping, followed by random rotations, as illustrated in Figure S4(a). The training loss curves, presented in Figure S4(b), demonstrate the convergence of the generator, discriminator, and high-frequency discriminator.

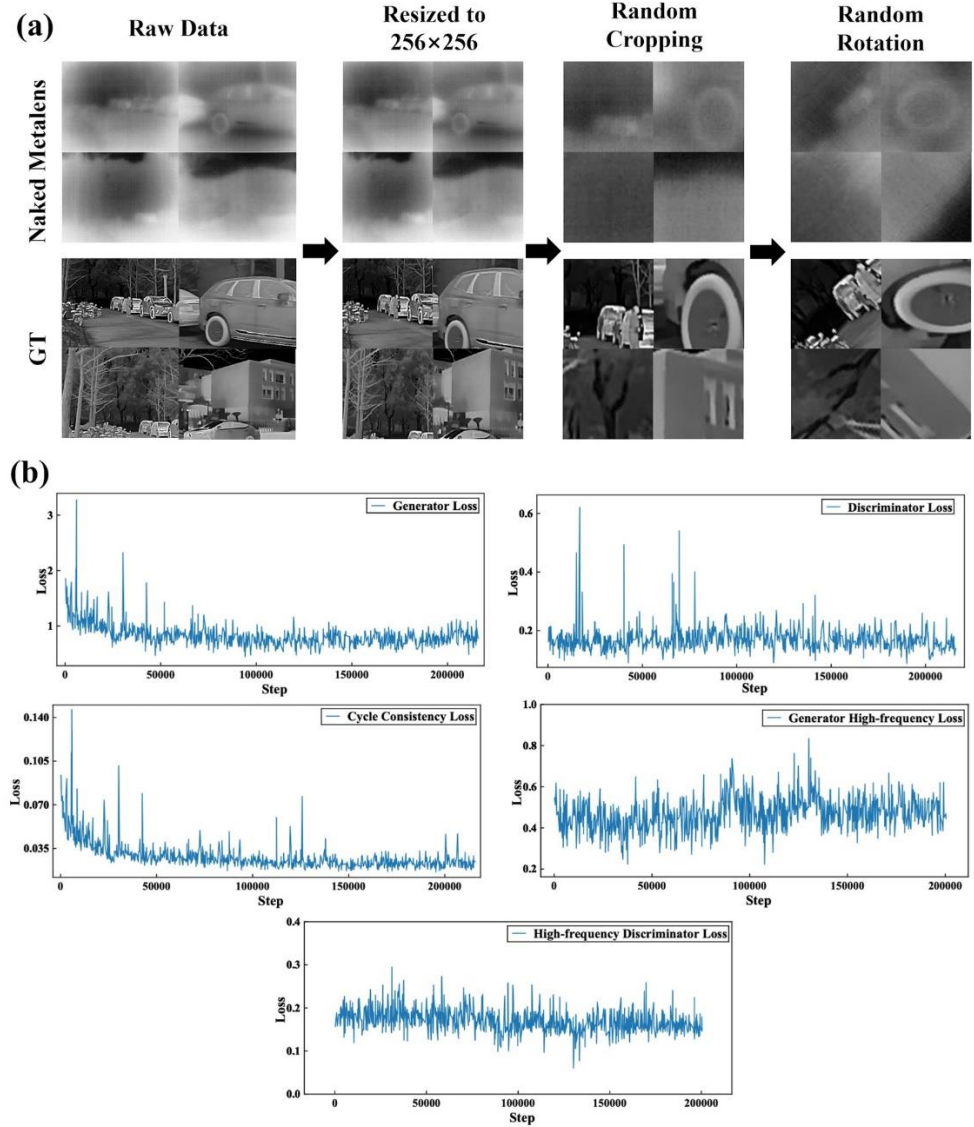

Fig. S4. Dataset preprocessing and training details. (a) The preprocessing steps for naked metalens videos and ground truth (GT) videos include resizing, random cropping, and rotations. (b) The training loss of the final model encompasses the losses of the generator, discriminator, and high-frequency discriminator.

**Evaluation Results of Test Videos**

We provided five test videos in total: Video1~5. The quantitative evaluation of Videos 1 to 3 is presented in the manuscript, while the evaluation for Videos 4 and 5 is displayed in Figure S3. For videos 4 and 5, our NNE Metalens achieves improvements of 4.67% and 4.13% in PSNR, and 4.69% and 2.90% in SSIM compared to the Mocycle-GAN, respectively. In terms of the FID metric, our method demonstrates even more substantial improvements, with percentages of 26.83% and 31.40% for Videos 4 and 5.

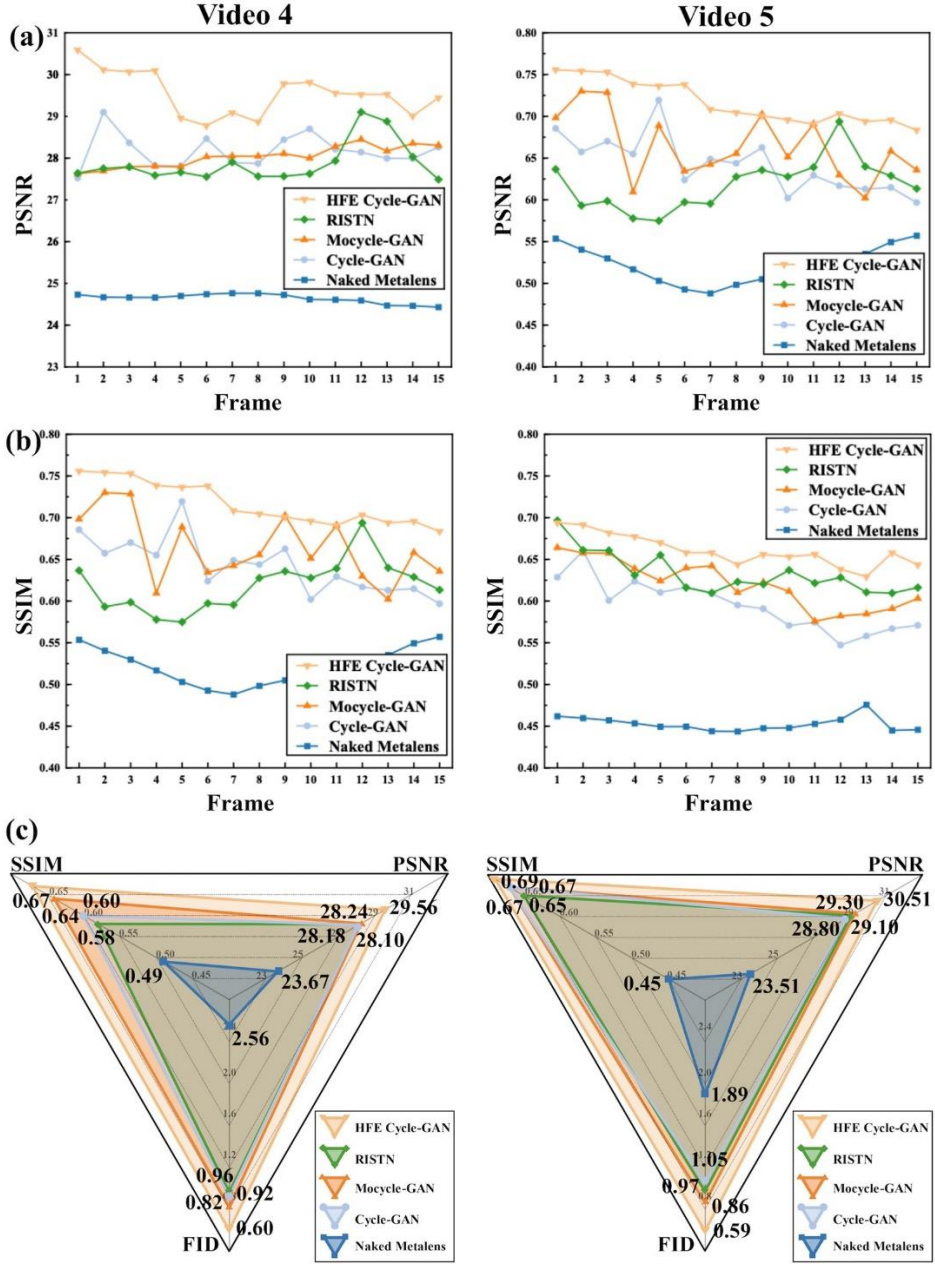

Fig. S5. Restoration evaluation of Videos 4 and 5 using PSNR, SSIM, and FID metrics. (a) and (b) present the PSNR and SSIM values, respectively, for 15 consecutive frames from each video. (c) shows the average PSNR, SSIM, and FID values for Videos 4 and 5. The NNE Metalens achieved improvements of 4.67% and 4.13% in PSNR, as well as 4.69% and 2.90% in SSIM for Videos 4 and 5, respectively, when compared to Mocycle-GAN. FID improvements of 26.83% and 31.40% are observed, indicating enhanced video quality and better feature preservation under low-contrast and noisy conditions.